\documentclass[aps,twocolumn,superscriptaddress,showpacs]{revtex4}
\usepackage{graphicx}%
\usepackage{dcolumn}
\usepackage{amsmath}

  \newcommand {\sio}             { \rm{SiO_{2}} }
  \newcommand {\na}              { \, ^{22}\rm{Na}  }
  \newcommand {\ci}              { \mu \rm{Ci}  }
  \newcommand {\nsec}            { \rm{ns}  }
  \newcommand {\kev}             { \rm{keV}  }
  \newcommand {\ev}              { \rm{eV}  }
  \newcommand {\mm}              { \rm{mm}  }
  \newcommand {\micron}          { \mu \rm{m}  }
  \newcommand {\cm}              { \rm{cm}  }
  \newcommand {\cc}              { \rm{g \, / cm}^{3}  }
  \newcommand {\doc}             { ^{\circ} \rm{C}  }
  \newcommand {\usecinverse}     { \mu \rm{s} ^{-1} }

  \newcommand {\Zeff}            { ^{1}Z_{{\rm eff}} }

\begin{document}

\bibliographystyle{apsrev}
  \title{
    Pickoff and spin-conversion quenchings of ortho-positronium in oxygen
  }

  \author{
    N.~Shinohara
  }
  \email{shino@rsaixsv.icepp.s.u-tokyo.ac.jp}

  \affiliation{
    Institute of Physics, College of Arts and Sciences, University of Tokyo, 3-8-1 Komaba, Meguro, Tokyo 153-8902, Japan
  }
  \author{
    T.~Chang
  }
  \affiliation{
    Institute of High Energy Physics, Academia Sinica, P.O. Box 2732, Beijing 100080, PR China  }
  \author{
  N.~Suzuki
  }
  \affiliation{
    The Institute of Physical and Chemical Research (RIKEN), Hirosawa 2-1, Wako, Saitama 351-0198, Japan
  }
  \author{
    T.~Hyodo
  }
  \affiliation{
    Institute of Physics, College of Arts and Sciences, University of Tokyo, 3-8-1 Komaba, Meguro, Tokyo 153-8902, Japan
  }
  \date{\today}


\begin{abstract}
The quenching processes of the thermalized ortho-positronium(o-Ps) on an oxygen molecule have been studied by the positron annihilation age-momentum correlation techinique(AMOC). The Doppler broadening spectrum of the 511 $\kev$ $\gamma$-rays from the 2$\gamma$ annihilation of o-Ps in ${\rm O}_{2}$ has been measured as a function of the o-Ps age. The rate of the quenching, consisting of the pickoff and the spin-conversion, is estimated from the positron lifetime spectrum. The ratio of the pickoff quenching rate to the spin-conversion rate is deduced from the Doppler broadening of the 511 $\kev$ $\gamma$-rays from the annihilation of the o-Ps. The pickoff parameter $\Zeff$, the effective number of the electrons per molecule which contribute to the pickoff quenching, for ${\rm O}_{2}$ is determined to be $0.6 \pm 0.4$. The cross-section for the elastic spin-conversion quenching is determined to be $(1.16 \pm 0.01) \times 10^{-19} \cm ^{2}$.
\end{abstract}


  \maketitle



  \section{Introduction}
    \label{sec:intro}
  Recently the annihilation of the low energy positrons on collision with various gas molecules has been studied systematically by Surko and his collaborators~\cite{Iwata95,Iwata00}. They have measured the annihilation rate, $\lambda_{+}$, of the thermalized positron and estimated the positron annihilation parameter, $Z_{{\rm eff}}$, defined as
\begin{equation}
  \lambda_{+}=\pi r_{0}^{2} c n Z_{{\rm eff}}\, ,
  \label{eq:Zeff}
\end{equation}
where $r_{0}$ is the classical electron radius, $c$ is the speed of light, and $n$ is the number density of the molecules. The $Z_{{\rm eff}}$ has been revealed to be very sensitive to small changes in the molecular structure and increasing parametrically from 1 to $10^{7}$~\cite{Iwata00}.

  A parameter similar to $Z_{{\rm eff}}$ is defined for the case of the positronium(Ps) annihilation in gases. It is related to the pickoff quenching of the ortho-positronium(o-Ps) and called $\Zeff$. This parameter represents the effective number of the electrons per molecule in a spin singlet state relative to the positron in the o-Ps~\cite{Fras68,Grif78,Char85}. The pickoff quenching rate is expressed as
\begin{equation}
  \lambda_{{\rm pickoff}}=4 \pi r_{0}^{2} c n \Zeff \, .
  \label{eq:1Zeff}
\end{equation}
  The values of $\Zeff$ for various gases reported so far are collected in Ref.~\cite{Char85}. They lie, except for ${\rm O}_{2}$, between 0.1 and 1.3, very small in contrast to $Z_{{\rm eff}}$.

In a paramagnetic gas such as ${\rm O}_{2}$, o-Ps can be also quenched by spin-conversion, {\itshape i.e.}, the conversion of o-Ps into para-positronium(p-Ps) followed by its prompt self-annihilation~\cite{Ferr58,Deut65,Kaki87-90}. There exists two kinds of Ps spin-conversion processes in ${\rm O}_{2}$. One excites the ${\rm O}_{2}$ molecule to the excited state ${\rm a}^{1}\Delta _{{\rm g}}$ (or ${\rm b}^{1}\Sigma_{{\rm g}}^{+}$), and thus may be called inelastic conversion. This process is active when the energy of the Ps is larger than $0.977 \ev$ (or $1.62 \ev$). The other leaves the ${\rm O}_{2}$ molecule in the ground state $X^{3}\Sigma_{{\rm g}}^{-}$ and may be called elastic conversion. The latter process has no threshold and thus active for the thermalized o-Ps. These two spin-conversions have quite different cross-sections~\cite{Deut65,Kaki87-90}; the cross-section for the former is on the order of $10^{-16} \cm^{2}$, and that for the latter is on the order of $10^{-19} \cm^{2}$~\cite{Kaki87-90}.

  In Ref.~\cite{Char85}, the quenching rate of o-Ps in ${\rm O}_{2}$ including the effect of the spin-conversion is expressed in terms of $\Zeff$ as $44 \pm 3$. The proper $\Zeff$ defined by Eq.~(\ref{eq:1Zeff}) is not known. In order to measure this, the pickoff quenching has to be separated from the spin-conversion.

  In the present work, we obtain the $\Zeff$ and the spin-conversion cross-section for ${\rm O}_{2}$ by positron annihilation age-momentum correlation technique(AMOC). The AMOC consists in a correlated measurement of the positron lifetime and the energy of the annihilation $\gamma$-rays. The Doppler broadening spectrum of the 511 $\kev$ $\gamma$-rays from the 2$\gamma$ annihilation of the o-Ps in ${\rm O}_{2}$ has been measured as a function of the time that the o-Ps atoms have spent from their formation to annihilation. The 2$\gamma$ annihilation in the time range in which only o-Ps exists results either from the pickoff quenching or from the self-annihilation of the p-Ps created through the spin-conversion. The $\Zeff$ and the spin-conversion cross-section are determined by separating the pickoff quenching from the spin-conversion.

  Preliminary results with a different setup were reported in Ref.~\cite{Chan92}.

  \section{Experimental}
    \label{sec:exp}

    \begin{figure}[htbp]
      \begin{center}
        \includegraphics{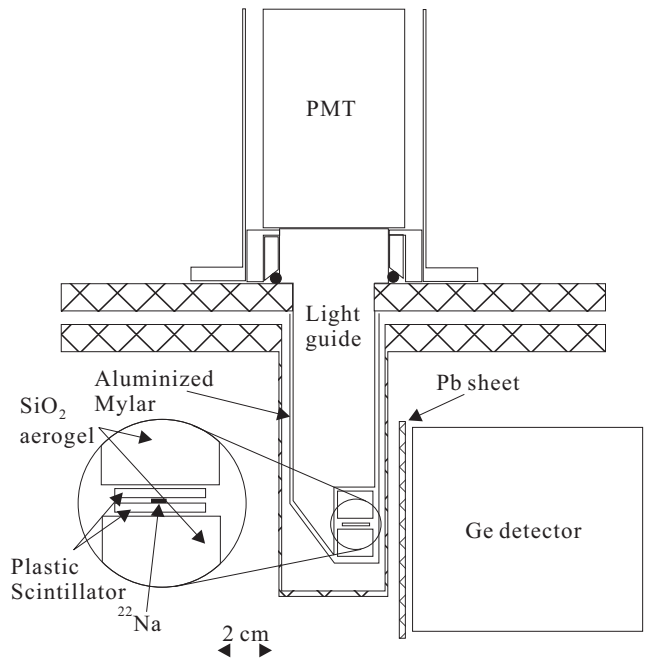}
        \caption{Schematic diagram of the experimental setup.}
        \label{fig:fig1}
      \end{center}
    \end{figure}

  The experimental setup is shown schematically in Fig.~\ref{fig:fig1}. A 0.5 $\ci$ $\na$ positron source was sandwiched between two sheets of 100 $\micron$ thick plastic scintillator. The source-scintillator assembly was placed between two pieces of 1 $\cm$ thick $\sio$ aerogel, which was used as Ps formation medium~\cite{Chan82}. The macroscopic density and the averaged grain diameter of the $\sio$ aerogel are 0.1 $\cc$ and about 5 ${\rm nm}$, respectively.

  Most of the positrons from the source pass through the scintillators and give scintillation lights. The lights pass through the transparent $\sio$ aerogel and are directed to a photomultiplier tube(PMT) by a light guide.
  A $\gamma$-ray emitted when a positron annihilates was detected by a high purity Ge detector. The output of the Ge detector was fed to a timing-filter-amplifier(TFA). The time interval between the anode signal of the PMT and the output of the TFA was converted into an peak amplitude of an output pulse by a time-to-amplitude converter(TAC) and used for the lifetime spectroscopy. The other output of the Ge detector was processed by a shaping amplifier and the peak hight was recorded for the energy information. The observed energy region was limited in the neighborhood of the 511 $\kev$ peak by a biased amplifier. All the data were stored in the list mode.

  A 2 $\mm$ thick ${\rm Pb}$ sheet was placed in front of the Ge detector to prevent low energy scattered $\gamma$-rays from simultaneously hitting the Ge detector.

  The time resolution was 4.8 $\nsec$(FWHM), and the energy resolution was 1.12 $\kev$(FWHM) at 512 $\kev$($\, ^{106}{\rm Ru}$).

  The chamber was filled with ${\rm O}_{2}$ of purity 99.9995$\%$ to 1.05 ${\rm atm}$ after evacuating it with a turbo molecular pump, and then isolated from the rest of the gas handling system. The measurement lasted for 16 hours. During the measurement, the room temperature was controlled to be $25.5 \pm 0.5 \doc$. For data analysis, a measurement without a gas was also made for 16 hours.

  \section{Results and discussion}
    \label{sec:res_and_dis}
      \begin{figure}[htbp]
        \begin{center}
          \includegraphics{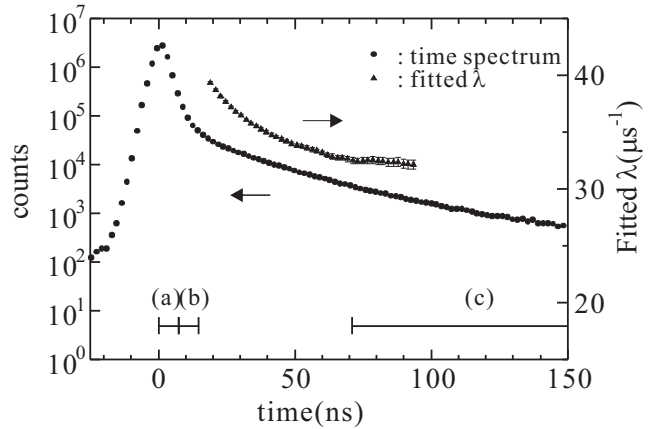}
          \caption{The time spectrum for the energy range from 500 $\kev$ to 516 $\kev$. The closed circles show the time interval distribution between a positron emission and the eventual annihilation. The closed triangles show the fittd $\lambda$ as a function of the start point of the fit. The energy spectra of the $511 \kev \gamma$-rays in the time windows (a), (b), and (c) are shown in Fig.~\ref{fig:fig4}. }
          \label{fig:fig2}
        \end{center}
      \end{figure}

  The closed circles in Fig.~\ref{fig:fig2} show the positron lifetime spectrum for the energy range from 500 $\kev$ to 516 $\kev$. The prompt peak is followed by the slow decay curve and subsequently the flat background. The prompt peak arises from the positron annihilations without Ps formation as well as the self-annihilations of p-Ps. The slow decay part originates from the o-Ps annihilations.
  The total annihilation rate of the o-Ps, $\lambda_{{\rm oPs}}$, is obtained from the slow decay part. It is the sum of the self-annihilation rate, $\lambda_{3\gamma}$, and the annihilation rates for three different collisional quenching modes;
\begin{equation}
  \lambda_{{\rm oPs}}=\lambda_{3\gamma}+\lambda_{\sio}+\lambda_{{\rm ox}}+\lambda_{{\rm spin}}\, ,
\end{equation}
where $\lambda_{\sio}$ is the rate of the pickoff quenching on the grain surface, $\lambda_{{\rm ox}}$ is that on the ${\rm O}_{2}$ molecule, and $\lambda_{{\rm spin}}$ is the rate of the spin-conversion on the ${\rm O}_{2}$ molecule.

  The latter two collisional quenching rates are propotional to the number density of the ${\rm O}_{2}$ molecules, $n$,
\begin{equation}
  \lambda _{{\rm q}}= n \sigma _{{\rm q}}v_{{\rm Ps}} \, ,
\label{eqn:lq}
\end{equation}
where $\sigma_{{\rm q}}$ is the quenching cross-section and $v_{{\rm Ps}}$ is the speed of the Ps. The speed of the o-Ps relative to the target molecule has been replaced by $v_{{\rm Ps}}$, neglecting the relatively small velocity of the molecules.

  The initial kinetic energy of the Ps depends on whether it is formed inside a $\sio$ grain or on a grain surface~\cite{Naga98}. The Ps formed and thermalized inside a $\sio$ grain escapes into the free space between the grains with the kinetic energy of 1 $\ev$ determined by the negative work function for the Ps. On the other hand, the Ps formed on the surfaces of the grains is emmitted with the kinetic energy of 3 $\ev$. It penetrates shallowly back into another grain and thermalizes, and reemitted with the kinetic energy of 1 $\ev$.

  Once the Ps energy reaches 1 $\ev$, it does not enter another grain any more. Then the energy loss of the Ps becomes much slower~\cite{Chan87,Naga95,Naga98b}. This leads to the time dependence of the collisional quenching rates, resulting in time dependence of the total annihilation rate $\lambda_{{\rm oPs}}$ and subtle and complicated shape of the o-Ps time spectrum. The time spectrum cannot be represented as a simple exponential decay before the o-Ps is thermalized. Denoting the survival probability of the o-Ps up to time t by $D(t)=\exp(-\int^{t}_{0} dt'\lambda_{{\rm oPs}}(t'))$, the time spectrum is given by
\begin{eqnarray}
  P(t) &=& N_{0}(-\epsilon \frac{d}{dt}D(t)) +C \nonumber \\
       &=& N_{0} [ \epsilon_{3\gamma}\lambda_{3\gamma}+\epsilon_{2\gamma} ( \lambda_{\sio}(t)+\lambda_{{\rm ox}}(t)\nonumber \\
       & & +\lambda_{{\rm spin}}(t) ) ] D(t) +C\, ,
\end{eqnarray}
 where $\epsilon_{3\gamma}$ and $\epsilon_{2\gamma}$ are the absolute detection efficiencies for the $\gamma$-rays from the 3$\gamma$ annihilations and the 2$\gamma$ annihilations, respectively, and $N_{0}$ and $C$ are constants.

  In order to determine the total annihilation rate $\lambda_{{\rm oPs}}$ for the themalized o-Ps, we fit the lifetime spectrum(Fig.~\ref{fig:fig2}) to a function A $\exp(-\lambda t)$ + B, where A and B are constants, and the parameter $\lambda$ is obtained as the start time of the fit $t^{\ast}$ is stepped out. The results are plotted by triangles as a function of the start time of the fit in Fig.\ref{fig:fig2}. The fitted $\lambda$ gradually decreases, indicating the o-Ps slowing down process, and then becomes flat, indicating that the o-Ps is thermalized. Once the o-Ps is thermalized, the fitted $\lambda$ represents the total annihilation rate of the o-Ps averaged over the Maxwell-Boltzmann velocity distribution at the measuring temperature. The $\lambda_{{\rm oPs}}$ for the thermalized o-Ps is determined from the value in the region where the fitted values are statistically consistent. We choose $t^{\ast} = 71 \nsec$ to yield
\begin{eqnarray}
  \lambda _{{\rm oPs}}&=&\lambda _{3\gamma}+\lambda _{\sio}^{\ast}+\lambda _{{\rm ox}}^{\ast}+\lambda_{{\rm spin}}^{\ast} \nonumber \\
                &=&32.5 \pm 0.2 \usecinverse\, ,
\end{eqnarray}
where the quenching rates for the thermalized o-Ps are represented with the superscript $^{\ast}$. The value for $\lambda_{3\gamma}+\lambda _{\sio}^{\ast}$ is determined similarly from the measurement without a gas, to be
\begin{eqnarray}
  \lambda_{{\rm oPs}}&=&\lambda_{3\gamma}+\lambda _{\sio}^{\ast} \nonumber \\
               &=&7.41 \pm 0.04 \usecinverse.
\end{eqnarray}
   By using $\lambda _{3\gamma}=7.040 \pm0.003\usecinverse$~\cite{Asai95}, the values for $\lambda_{\sio}^{\ast}$ and $\lambda_{{\rm ox}}^{\ast}+\lambda_{{\rm spin}}^{\ast}$ are determined to be,
\begin{eqnarray}
  \lambda_{\sio}^{\ast}&=&0.37 \pm 0.04 \usecinverse
    \label{eq:lsio} \\
  \lambda_{{\rm ox}}^{\ast}+\lambda_{{\rm spin}}^{\ast}&=&25.1 \pm 0.2 \usecinverse.
\end{eqnarray}


      \begin{figure}[htbp]
        \begin{center}
          \includegraphics{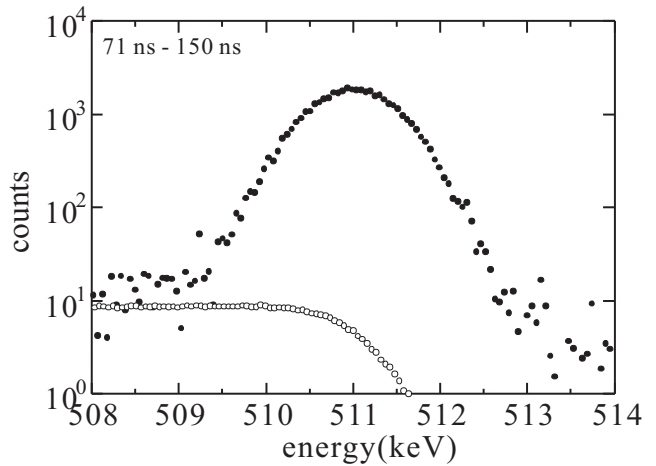}
          \caption{The energy spectrum of the annihilation $\gamma$-rays(closed circles) for 1 ${\rm atm}$ of ${\rm O}_{2}$ with the time range from 71 ${\rm ns}$ to 150 ${\rm ns}$. The background has been subtracted. The open circles show the o-Ps self-annihilation component.}
          \label{fig:fig3}
        \end{center}
      \end{figure}

  Figure~\ref{fig:fig3} shows an example of the time-selected energy spectrum for 1 ${\rm atm}$ of ${\rm O}_{2}$ measured by the AMOC. The intensity and the shape of the background spectrum have been estimated from the energy spectrum in the time range from 400 $\nsec$ to 700 $\nsec$ and subtracted.


      \begin{figure}[htbp]
        \begin{center}
          \includegraphics{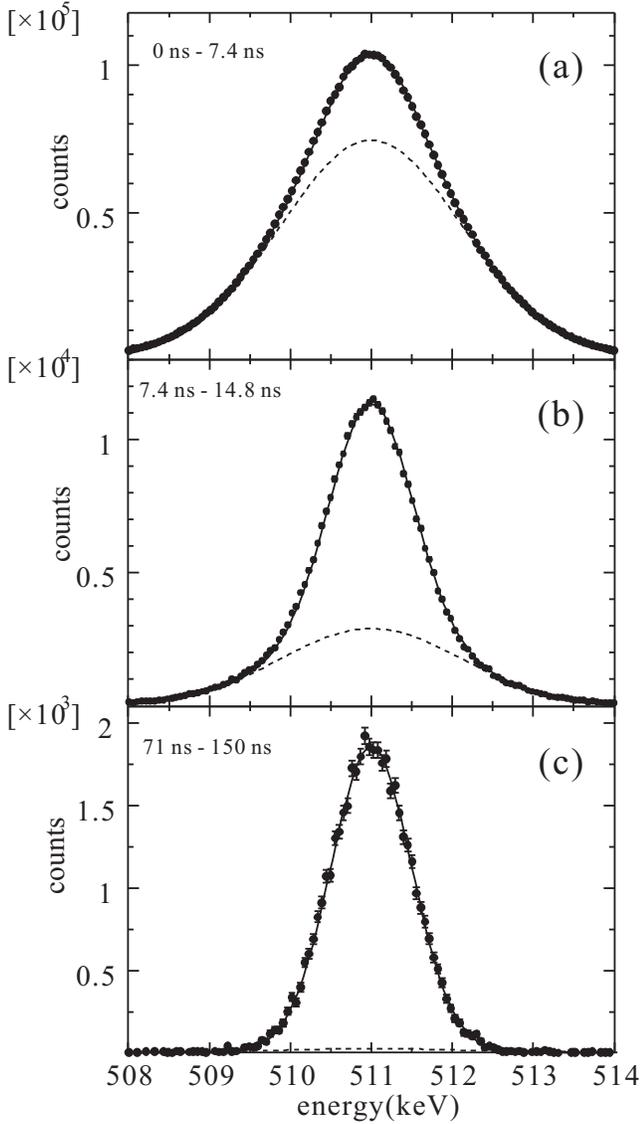}
          \caption{Examples of the time-selected energy spectra of the 511$\kev$ $\gamma$-rays for 1 ${\rm atm}$ of ${\rm O}_{2}$. }
          \label{fig:fig4}
        \end{center}
      \end{figure}

  Figure~\ref{fig:fig4} shows the Doppler broadening spectra in the time ranges indicated. The spectra have been corrected for the 3$\gamma$ annihilation component, which is estimated using theoretical 3$\gamma$ spectrum convoluted with the energy resolution curve, {\itshape i.e.}, the shape of the 512 $\kev$ peak from $\, ^{106}\rm{Ru}$ and normalized to the counts in the energy range from 485 $\kev$ to 500 $\kev$, as shown in Fig.~\ref{fig:fig3}.

  The spectra in Fig.\ref{fig:fig4} are fitted to two gaussian functions(solid curves). The dashed curves show the broad component. The width of this component was fixed to that of the pickoff component for $\sio$ aerogel only, because the change in the width due to the presence of ${\rm O}_{2}$ molecules was not appreciable\cite{Kaki87-90}.

\begin{figure}[htbp]
        \begin{center}
          \includegraphics{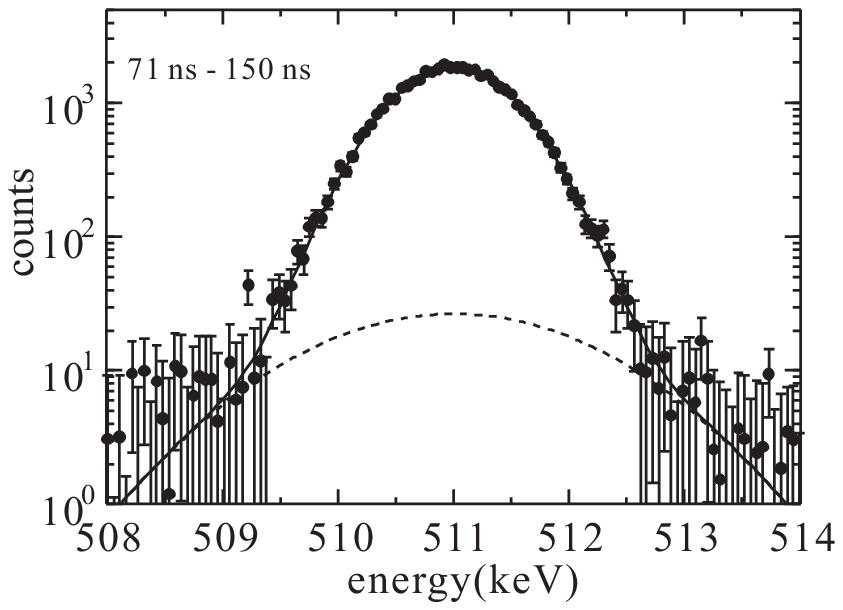}
          \caption{Doppler broadening spectrum of the annihilation $\gamma$-rays from the thermalized o-Ps. The chance coincidence and the $3\gamma$ backgrounds have been subtracted. The solid curve shows the two component gaussian fit to the data. The dashed curve shows the fitted pickoff quenching component and the rest is the spin-conversion component.}
          \label{fig:fig5}
        \end{center}
      \end{figure}

  The spectra in Fig.~\ref{fig:fig4}(a) and (b) include the components which contributes to the prompt peak; {\itshape i.e.}, those from the annihilations of the non-Ps positrons and p-Ps. Hence they are not appropriate to the analysis. The spectrum in Fig.~\ref{fig:fig4}(c) represents the $2\gamma$ annihilations of the thermalized o-Ps. The pickoff quenching gives the extremely low intensity broad component representing the momentum distribution of the electrons bound in ${\rm O}_{2}$ molecules and those on the $\sio$ surfaces. The narrow component results from the spin-conversion quenching and represents the center-of-mass momentum distribution of the p-Ps at the moment of the annihilation after the conversion from o-Ps. A log scale is used for the vertical axis in Fig.~\ref{fig:fig5} to blow up the pickoff quenching component.

  The ratio of the intensity of the broad component to that of the narrow component is
\begin{equation}
  \frac{I_{{\rm broad}}}{I_{{\rm narrow}}}=(3.3 \pm 1.3) \times 10^{-2} .
\label{eq:ValueofRatio}
\end{equation}
  An alternative analysis with the width of the broad component as a free parameter gives a similar result.
  The intensity of the 2$\gamma$ quenching component in the time-selected energy spectrum, $I_{{\rm q}}$(q representing $\sio$, ${\rm ox}$, ${\rm spin}$), for the time range from $t^{\ast}$ to 150 $\nsec$ is given by
\begin{eqnarray}
  I_{{\rm q}}&=&\int^{150\nsec}_{t^{\ast}} N_{0} \epsilon_{2\gamma}\lambda_{{\rm q}}(t)D(t) dt \nonumber \\
             &=&\lambda_{{\rm q}}^{\ast}\int^{150\nsec}_{t^{\ast}} N_{0} \epsilon_{2\gamma}D(t) dt \,.
\end{eqnarray}
Hence we have the relation:
\begin{equation}
  \frac{I_{{\rm broad}}}{I_{{\rm narrow}}} =\frac{I_{\sio}+I_{{\rm ox}}}{I_{{\rm spin}}}
                               =\frac{\lambda_{\sio}^{\ast}+\lambda_{{\rm ox}}^{\ast}}{\lambda_{{\rm spin}}^{\ast}} \,.
\label{eq:ratio}
\end{equation}

Combining the results (\ref{eq:lsio}) $\sim$ (\ref{eq:ValueofRatio}), and (\ref{eq:ratio}), the quenching rates for ${\rm O}_{2}$ are obtained as,
\begin{eqnarray}
  {\rm pickoff\,\,quenching\,\,rate}&:& \lambda_{{\rm ox}}^{\ast}=0.4\pm0.3\usecinverse\\
  {\rm spin\,\,conversion\,\,rate}&:& \lambda_{{\rm spin}}^{\ast}=24.7\pm0.2\usecinverse .
\end{eqnarray}
From the relation(\ref{eq:1Zeff}), we conclude that
\begin{eqnarray*}
  \Zeff=0.6 \pm 0.4 \,.
\end{eqnarray*}
This value is on the order of magnitude as the other gases~\cite{Char85}.

  The spin-conversion cross-section $\sigma_{{\rm spin}}$ is
\begin{eqnarray*}
  \sigma_{{\rm spin}}=(1.16 \pm 0.01) \times 10^{-19} \cm ^{2} \,.
\end{eqnarray*}
This cross-section is for the elastic conversion process~\cite{Kaki87-90}.

\vspace{1em}

    \label{sec:conc}
In conclusion, we have studied the 2$\gamma$ annihilation of the thermalized o-Ps in oxygen by AMOC. The $\Zeff$ and the elastic spin-conversion cross-section of the thermalized o-Ps are estimated by separating the pickoff quenching from the spin-conversion. The $\Zeff$ for ${\rm O}_{2}$ is revealed to be on the order of magnitude as the other gases.

  \section*{Acknowledgement}
We would like to acknowledge Dr. Y. Nagashima and Dr. H. Saito for valuable discussions.


\end{document}